\documentclass[aps,twocolumn,superscriptaddress,footinbib,floatfix,showpacs]{revtex4-1}
\usepackage{graphicx}
\usepackage{comment}
\usepackage{MnSymbol,color}
\usepackage{tikz}
\usepackage{pgfplots}
\usepackage[export]{adjustbox}
\usetikzlibrary{plotmarks}
\definecolor{bgreen}{rgb}{0.0,0.5,0.0}
\definecolor{bblue}{rgb}{0.0,0.0,0.9}
\definecolor{bgold}{rgb}{0.7,0.5,0.0}
\definecolor{bred}{rgb}{0.9,0.0,0.0}

\newcommand{\ba}{\mathbf{a}}
\newcommand{\br}{\mathbf{r}}
\newcommand{\bx}{\mathbf{x}}
\newcommand{\bv}{\mathbf{v}}

\newcommand{\bD}{\mathbf{D}}
\newcommand{\myS}{\mathcal{S}1}

\begin{document}

\title{\textcolor{black}{Submesoscale dispersion in the vicinity of the Deepwater Horizon spill} 
}

\author{Andrew C. Poje}
\affiliation{City University of New York, New York, USA}
\author{Tamay M. \"{O}zg\"{o}kmen}
\affiliation{University of Miami, Miami, USA}
\author{Bruce Lipphardt, Jr.}
\affiliation{University of Delaware, Newark, USA}
\author{Brian K. Haus}
\author{Edward H. Ryan}
\author{Angelique C.  Haza}
\author{A.J.H.M. Reniers}
\author{Josefina Olascoaga}
\author{Guillaume Novelli}
\author{Francisco J. Beron-Vera}
\author{Shuyi Chen}
\author{Arthur J. Mariano}
\affiliation{University of Miami, Miami, USA}
\author{Gregg Jacobs}
\author{Pat Hogan}
\author{Emanuel Coelho}
\affiliation{Naval Research Laboratory, Stennis Space Center, Mississippi, USA}
\author{A.D. Kirwan, Jr.}
\author{Helga Huntley}
\affiliation{University of Delaware, Newark, USA}
\author{Annalisa Griffa}
\affiliation{CNR, La Spezia, Italy}

\begin{abstract}
Reliable forecasts for the dispersion of oceanic contamination are important for coastal ecosystems, society and 
the economy as evidenced by the Deepwater Horizon oil spill in the Gulf of Mexico in 2010 and the Fukushima nuclear plant
incident in the Pacific Ocean in 2011.
Accurate prediction of pollutant pathways and concentrations at the ocean surface requires understanding
ocean dynamics over a broad range of spatial scales.
Fundamental questions concerning the structure of the velocity field at the submesoscales
(100 meters to tens of kilometers, hours to days) remain unresolved due to a lack of synoptic measurements
at these scales.
\textcolor{black}
{Using high-frequency position data provided by the near-simultaneous release of hundreds of
accurately tracked surface drifters, 
we study the structure of submesoscale surface velocity fluctuations
in the Northern Gulf Mexico.
Observed two-point statistics confirm the accuracy of
classic turbulence scaling laws at 200m$-$50km scales
and clearly indicate that dispersion at the submesoscales is \textit{local}, driven predominantly by
energetic submesoscale fluctuations.}
The results demonstrate the feasibility and utility of deploying large clusters of drifting instruments to provide 
synoptic observations of spatial variability of the ocean surface velocity field. Our findings
allow quantification of the submesoscale-driven dispersion missing in current operational circulation
models and satellite altimeter-derived velocity fields.
\end{abstract}

\maketitle

\section{Introduction}
The Deepwater Horizon (DwH) incident was the largest accidental oil spill into marine waters in history with some
4.4 million barrels released into the DeSoto Canton of the northern Gulf of Mexico (GoM) from a subsurface pipe 
over approximately 84 days in the Spring and Summer of 2010\textsuperscript{\cite{Crone2010}}.
Primary scientific questions, with immediate practical implications, arising from such catastrophic pollutant 
injection events are the path, speed and spreading rate of the pollutant patch.
Accurate prediction requires knowledge of the ocean flow field at all relevant temporal and spatial scales.
While ocean general circulation models were widely used during and after the DwH 
incident\textsuperscript{\cite{Mezic2010, Mariano2011, Olascoaga2012,
Maltrud2010,Huntley2011}},
such models only capture the main mesoscale processes (spatial scale
larger than 10 km) in the GoM. %Smaller scale processes must necessarily be parameterized.
The main factors controlling surface dispersion in the DeSoto Canyon region remain unclear. The region lies between
the mesoscale-eddy driven deep water GoM\textsuperscript{\cite{Olascoaga2013}} and the wind-driven shelf\textsuperscript{\cite{Morey2003}} 
while also being subject to the 
buoyancy input of the Mississippi River plume during the spring and summer months\textsuperscript{\cite{Walker2005}}. 
Images provided by the large amounts of surface oil produced in the  DwH incident  revealed a rich array of flow patterns\textsuperscript{\cite{Jones2011}}
showing organization of surface oil not only by mesoscale straining into the Loop Current Eddy Franklin, 
but also \textcolor{black}{by submesoscale processes}. Such
processes operate at spatial scales and involve physics not presently captured in operational circulation models.
Submesoscale motions, where they exist, can directly influence the local transport of
biogeochemical tracers\textsuperscript{\cite{klein2009,levy2012a}} and provide pathways
for energy transfer from the wind-forced mesoscales to the dissipative microscales\textsuperscript{\cite{Mac2008,foxkemper2011,Vallis2013}}.
Dynamics at the submesoscales have been subject of recent research
\textsuperscript{\cite{muller2005,capet2008,foxkemper2008,taylor2011,dasaro2011}}.
However, the investigation of their effect on ocean transport has been predominantly
modeling based\textsuperscript{\cite{Mac2008,poje2010,haza2012,ozgokmen12b}} and
synoptic observations, at adequate spatial and temporal resolutions, are rare
 \textsuperscript{\cite{bracco2012,andrey2013}}. The mechanisms responsible for the establishment,
 maintenance, and energetics of such features in the Gulf of Mexico remain
unclear.

Instantaneous measurement of all representative spatiotemporal scales 
of the ocean state is notoriously difficult\textsuperscript{\cite{Sanford2011}}. 
As previously reviewed\textsuperscript{\cite{ozgokmen12a}}, traditional observing systems are not ideal for synoptic
sampling of near-surface flows at the submesoscale. 
Owing to the large spacing between ground tracks\textsuperscript{\cite{Ducet2000}} and along-track signal 
contamination from high-frequency 
motions\textsuperscript{\cite{Chavanne2010}}, gridded altimeter-derived sea level anomalies only resolve the largest
submesoscale motions.
Long time-series ship-track current measurements attain similar, larger than 2 km, spatial resolutions, 
and require averaging the observations over evolving ocean states\textsuperscript{\cite{Callies2013}}.
Simultaneous, two-point ADCP measurements from pairs of
ships\textsuperscript{\cite{andrey2013}} provide sufficient resolution to
show the existence of energetic submesoscale fluctuations in the mixed-layer
but do not explicitly quantify the scale-dependent transport induced by such motions at the surface.
Lagrangian experiments, centered on tracking large numbers of water-following instruments, provide the most feasible means of 
obtaining spatially distributed, simultaneous 
measurements of the structure of the  ocean's surface velocity field on 100m to 10km length scales.

Denoting a trajectory by $\bx(\ba,t)$ where $\bx(\ba,t_0) = \ba$, the relative separation of a particle pair is given by $
\bD(t,\bD_0) =  \bx(\ba_1,t) - \bx(\ba_2,t)  = 
\bD_0 + \int_{t_0}^t \Delta \bv(t',\bD_0) dt'$
where the Lagrangian velocity difference is defined by 
$\Delta \bv(t,\bD_0) =  \bv(\ba_1,t) -  \bv(\ba_2,t)$.
The statistical quantities of interest, both practically and theoretically, 
are the scale-dependent relative dispersion $D^2(t)=\langle \bD \cdot \bD \rangle$ (averaged over particle pairs)
and the average longitudinal or separation velocity, 
$\Delta v(r)$, at a given separation, $r$. 
The velocity scale is defined by the second order structure function 
$\Delta v(r) = \sqrt{\langle \delta v ^2 \rangle}$ where 
$\delta v(r) = \left( \bv(\bx + \br) - \bv(\bx) \right) \cdot  \br/|| \br ||$
\textsuperscript{\cite{moninyaglom71,Babiano1990}} where the averaging is now conditioned on the pair separation $r$.

The applicability of classical dispersion theories\textsuperscript{\cite{Richardson:1926,Batchelor:1952,Babiano1990}} developed in the context of 
homogeneous, isotropic turbulence with localized spectral forcing, to ocean flows subject to
the effects of rotation, stratification and complex forcing at disparate length and time scales remains unresolved.
Turbulence theories broadly predict two distinct dispersion regimes depending upon the shape of
the spatial kinetic energy spectrum, $E(k) \sim k^{-\beta}$, of the velocity field\textsuperscript{\cite{Bennett:1984}}. 
For sufficiently steep spectra ($\beta \ge 3$)
the dispersion is expected to grow exponentially, $ D \sim \mathrm{e}^{\lambda t}$ with a 
scale-independent rate.
At the submesoscales ($\sim100\mathrm{m} - 10$km), this non-local growth rate  
will then be determined by 
the mesoscale motions currently resolved by predictive models.  
For shallower spectra ($1 < \beta < 3$), however, the dispersion is local,
$D \sim t^{2/(3-\beta)}$, and the growth rate of a pollutant patch is  
dominated by advective processes at the scale of patch. 
Accurate prediction of dispersion in this regime requires resolution of the advecting field at 
smaller scales than the mesoscale.

\begin{figure*}[ht]
\centerline{
\includegraphics[width=0.95\textwidth]{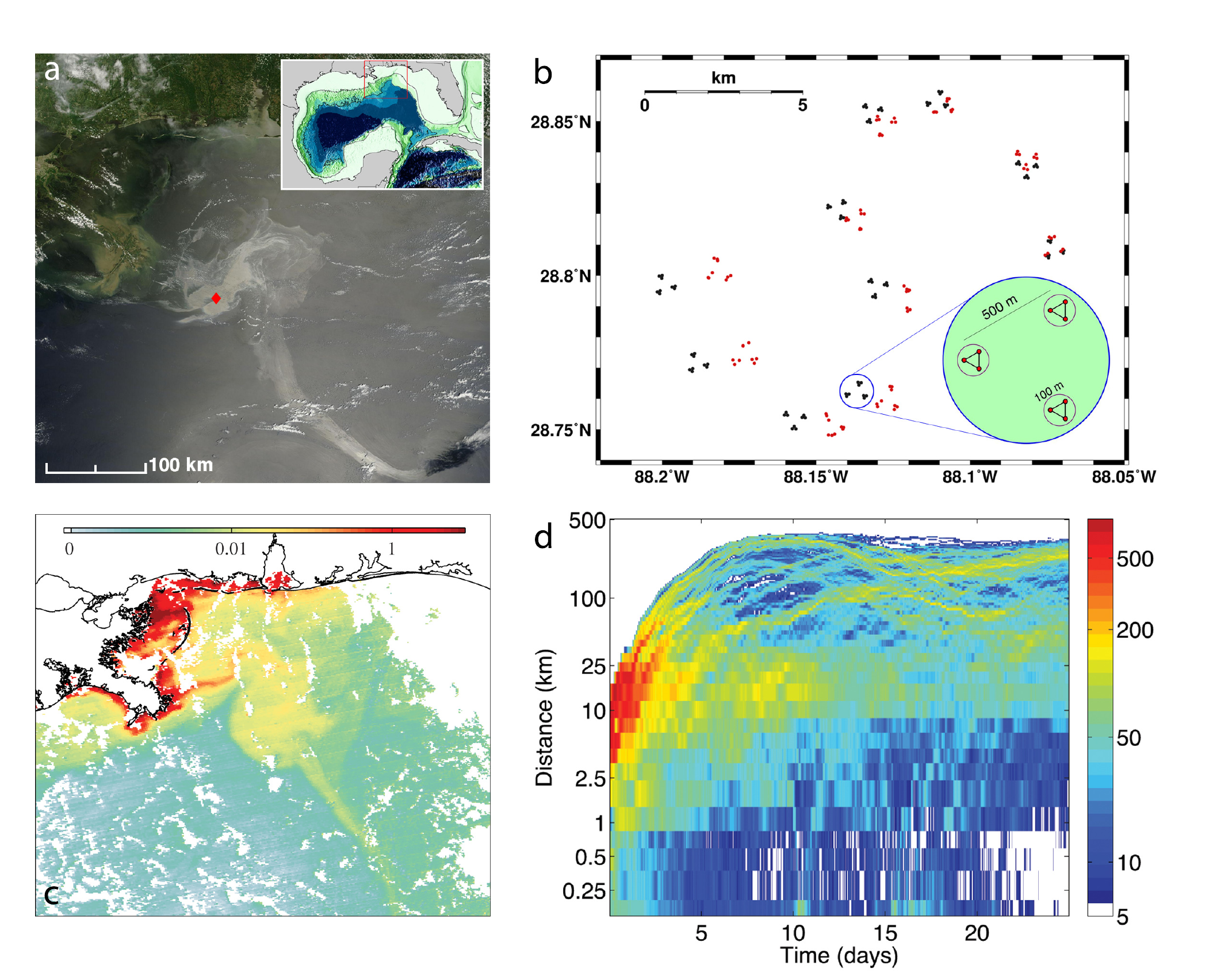}
}
\caption{{\bf Multi-scale flows near the DwH and DeSoto Canyon region.}
(a) Synthetic aperture radar (SAR) image of the DwH oil slick taken on 17 May 2010.
The red diamond marks the location of the DwH wellhead, and the inset shows the geographic location. 
(b) Drifter launch patterns: The actual pattern obtained (red circles) for $\mathcal{S}1$  at the launch time of the last drifter 
compared to the targeted template (black circles).  The inset shows a single node of the multi-scale launch
pattern.
%Animation of the actual release is available at: {\em https://vimeo.com/73830566}. 
(c) Chlorophyll-\emph{a} concentration \textcolor{black}{(indicative of phytoplankton suspended in the upper ocean flows)} derived from the MODIS sensor aboard 
the \emph{Aqua} satellite on 12 July 2012.  \textcolor{black}{The similarity of this image to (a) indicates 
that the GLAD experiment sampled flow conditions similar to those during the spill.}
(d) The time evolution of the number of drifter pairs at given separation distances for the $\mathcal{S}2$ release (pair numbers
on log-scale).}
\label{fig1}
\end{figure*}

While compilations of data from dye measurements broadly support local dispersion in natural flows\textsuperscript{\cite{Okubo1971}}, 
the range of scales in any particular dye experiment is limited. 
A number of Lagrangian observational studies have attempted to fill this gap.
LaCasce and Ohlmann\textsuperscript{\cite{LaCasce2003}} considered 140 pairs of surface drifters on 
the GoM shelf over a five year period and found evidence of a non-local regime for temporally smoothed data at 1-km scales.
Koszalka \textit{et al}\textsuperscript{\cite{Koszalka2009}} using $\mathcal{O}$(100) drifter pairs with $D_0 < 2$km
launched over ~18 months in the Norwegian Sea, found an exponential fit for $D^{2}(t)$ for a limited time ($t = 0.5-2$ days),
although the observed longitudinal velocity structure function is less clearly fit by a corresponding quadratic.
They concluded that a non-local dispersion regime could not be identified.
In contrast, Lumpkin and Elipot\textsuperscript{\cite{Lumpkin2010}} found evidence of local dispersion at 1km scales using 15-m drogued drifters
launched in the winter-time North Atlantic. It is not clear how the accuracy of the Argos positioning system 
(150-1000m) used in these studies affects the submesoscale dispersion estimates.
Schroeder \textit{et al}\textsuperscript{\cite{Schroeder2012}}, specifically targeting a coastal front using a multi-scale sampling pattern,
obtained results consistent with local dispersion, but the statistical significance (maximum 64 pairs) remained too low to be definitive.

\begin{figure*}[ht]
\begin{center} 
\includegraphics[width=0.45\textwidth]{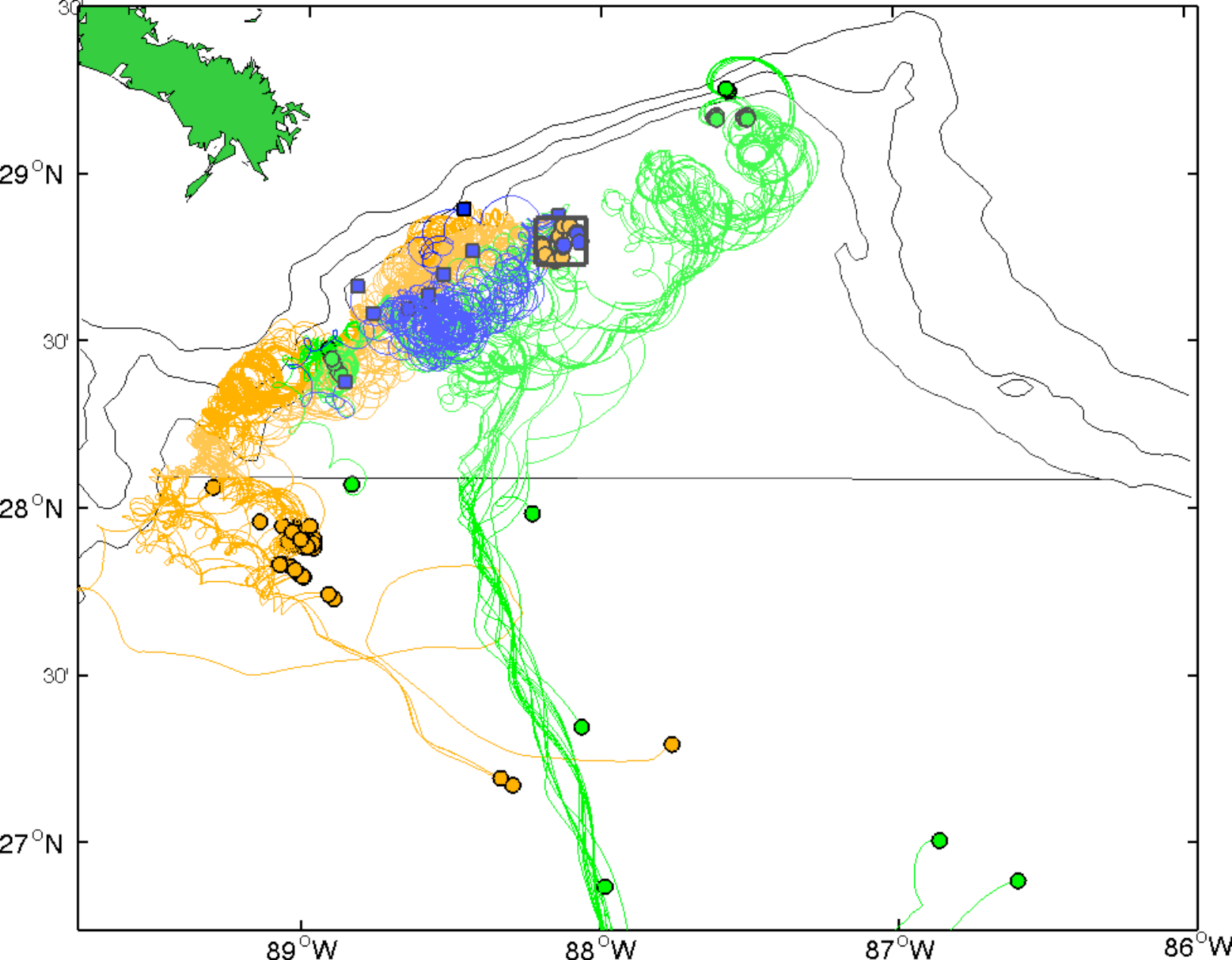}
\includegraphics[width=0.43\textwidth]{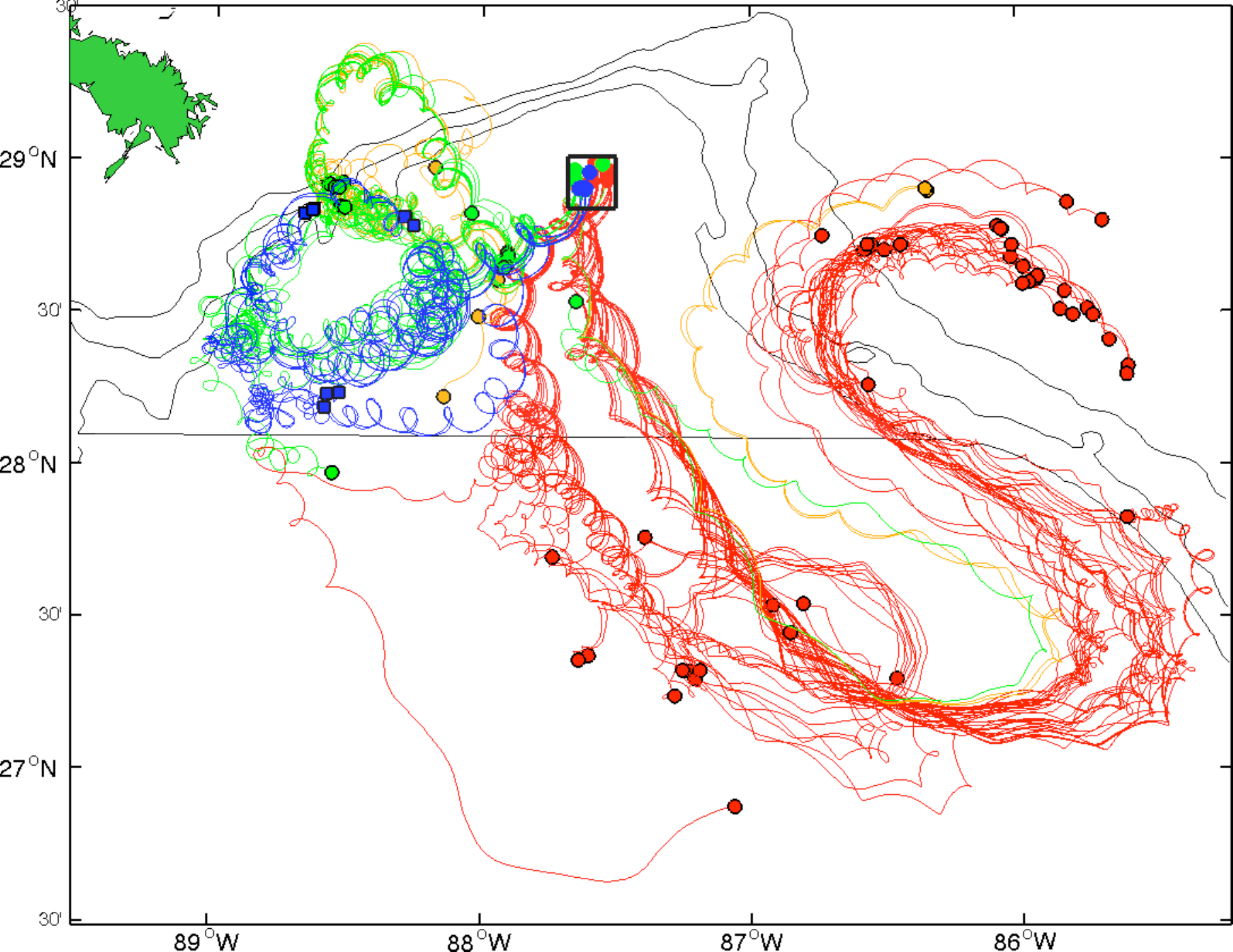}
\end{center}
\caption{{\bf GLAD trajectories.}
Trajectories for $\mathcal{S}$1 and $\mathcal{T}$1 (left panel) and $\mathcal{S}$2 (right panel) with initial and Day-21 positions marked by symbols. 
Trajectories
are color-coded based on total residence time, $\tau$,  in the canyon:
red triangles 
%\protect\tikz \protect\draw[mark=triangle*,mark size=5pt, mark options={color=black, solid, rotate=90, fill=bred }] plot coordinates {(0,0)};
for $\tau < 7$ days,
%\protect\tikz \protect\draw[mark=*,mark size=5pt, mark options={color=black, solid, fill=bgold }] plot coordinates {(0,0)};
gold circles
for $7 < \tau < 14$ days,
%\protect \tikz \protect\draw[mark=*,mark size=5pt, mark options={color=black, solid, fill=bgreen }] plot coordinates {(0,0)};
green circles
for  $ 14 < \tau < 21$ days and
% \protect \tikz \protect\draw[mark=square*,mark size=5pt, mark options={color=black, solid, fill=bblue }] plot coordinates {(0,0)};
and blue squares
for $\tau > 21$ days. The zonal line at 28.1$^{\circ}$N marks the latitude used as boundary for residence time estimates inside the Canyon.}
\label{fig2}
\end{figure*}

\section{Results}
The primary goal of the Grand Lagrangian Deployment (GLAD) experiment was to quantify the scale-dependence of the surface-velocity field from synoptic observations of two-point Lagrangian position and velocity increments
by simultaneously deploying an unprecedented number of  drifters. 
The critical program design element was the use of approximately 300 GPS-equipped CODE drifters\textsuperscript{\cite{Davis1985}} 
to provide a new level of statistical accuracy in measuring two-point Lagrangian velocity and displacement statistics.
\textcolor{black}{CODE drifters, with submerged sails approximately 1m deep by 1m wide,
are designed and tested to follow upper-ocean flows in the presence of wind and waves.}
All GLAD drifters were launched during the period of 20 July to 31 July 2012; during the same season as the DwH event two years earlier.
A satellite  sea-surface color image taken eight days before the first GLAD drifter launch shows striking similarities to satellite images 
during the DwH event (Figs.\ \ref{fig1}a,c).

To obtain high densities of multi-point,
contemporaneous position and velocity data at a range of separation scales spanning the meso-submesoscale boundary,
drifters were released in a space-filling $\mathcal{S}$ configuration
within an area approximately  8km $\times$ 10km. The configuration provides synoptic sampling at the upper boundary 
of the submesoscale range while minimizing the time to execute the deployment with a single ship. 
The $\mathcal{S}$ track consists of 10 nodes spaced at 2km 
with each node containing nine drifters arranged in triplets of nested equilateral triangles, with separations of 100m between drifters within a triplet and of 500m between triplets within a node (Fig. \ref{fig1}b). 
The pattern allows simultaneous sampling of multiple separation scales between 100m and 10km. 
The typical duration for the release of all 90 drifters was approximately five hours. 
The evolution of the number of particle pairs at given separation distances
(Fig.\ \ref{fig1}d), indicates that large numbers of simultaneous drifter pairs, especially at submesoscale separations, were 
obtained.

\begin{figure*}[ht]
\begin{center} 
\includegraphics[width=0.45\textwidth, valign=t]{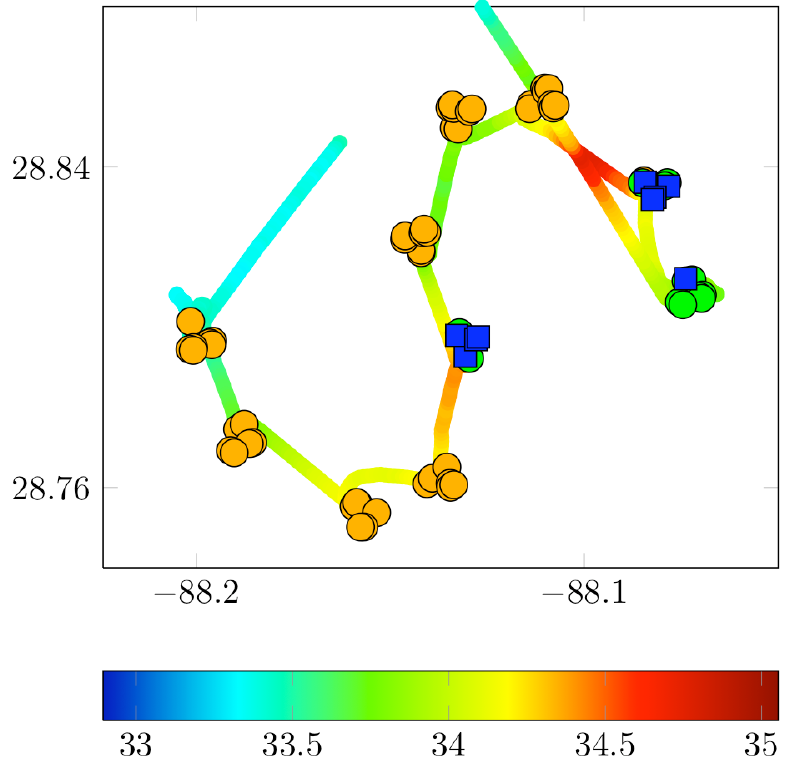} 
\includegraphics[width=0.43\textwidth, valign=t]{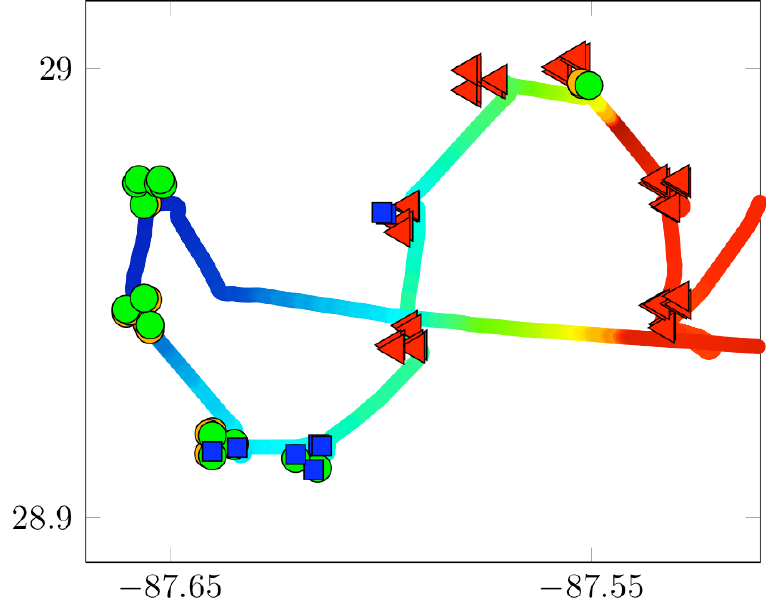}
\end{center}
\caption{{\bf Sensitivity to launch positions.} 
Initial launch locations and ship-track sea-surface salinity maps for $\mathcal{S}$1 (left panel) and $\mathcal{S}$2 (right panel) 
launches. Initial conditions
are color-coded based on total residence time in the canyon. Refer to the caption of Fig. \ref{fig2} for  the description of color coding. \textcolor{black}{The colored tracks and the color bar indicate sea surface salinity measured along ship track.}}
\label{fig6}
\end{figure*}

Initial 21-day trajectories for three drifter clusters launched  within the DeSoto Canyon, 
$\mathcal{S}$1 (near the DwH site, 89 drifters), $\mathcal{S}$2 (central  DeSoto Canyon targeting a surface salinity front, 90 drifters), 
and $\mathcal{T}$1 (northern tip of DeSoto Canyon, 27 drifters)
are shown in Fig.\ \ref{fig2}. 
The degree of confinement of surface water within the canyon and the role played by observed surface density fronts are
quantified by drifter residence time statistics. 
Trajectories in Fig.\ \ref{fig2} are coded by residence time, defined as the total amount of time spent within the closed region bounded by the 1,000-m  isobath and the 28.1$^\circ$-N latitude line over the 28-day period after launch.
The residence time for all drifters in the $\mathcal{S}$1 and $\mathcal{T}$1 deployments is longer than one week with a large number of 
$\mathcal{S}$1 drifters remaining
within the canyon for more than a month. 
Residence times for drifters in the $\mathcal{S}$2 launch, specifically those 
targeting a frontal feature in surface density, show much larger variation. 

Surface salinity measurements (Fig.\ \ref{fig6}) reveal a highly foliated horizontal-density structure associated with the
variability in the Mississippi river outflow (MRO) plume. 
Residence times in both $\mathcal{S}$ launches are extremely sensitive to launch location 
and are strongly correlated with initial salinity.
This is especially true for the $\mathcal{S}$2 launch where drifters launched on the 
more saline eastern side of the front rapidly exit the canyon while those launched in less saline water 
remain within the western canyon for considerably longer times.

\begin{figure*}[ht]
\begin{center}
\includegraphics[width=0.62\textwidth]{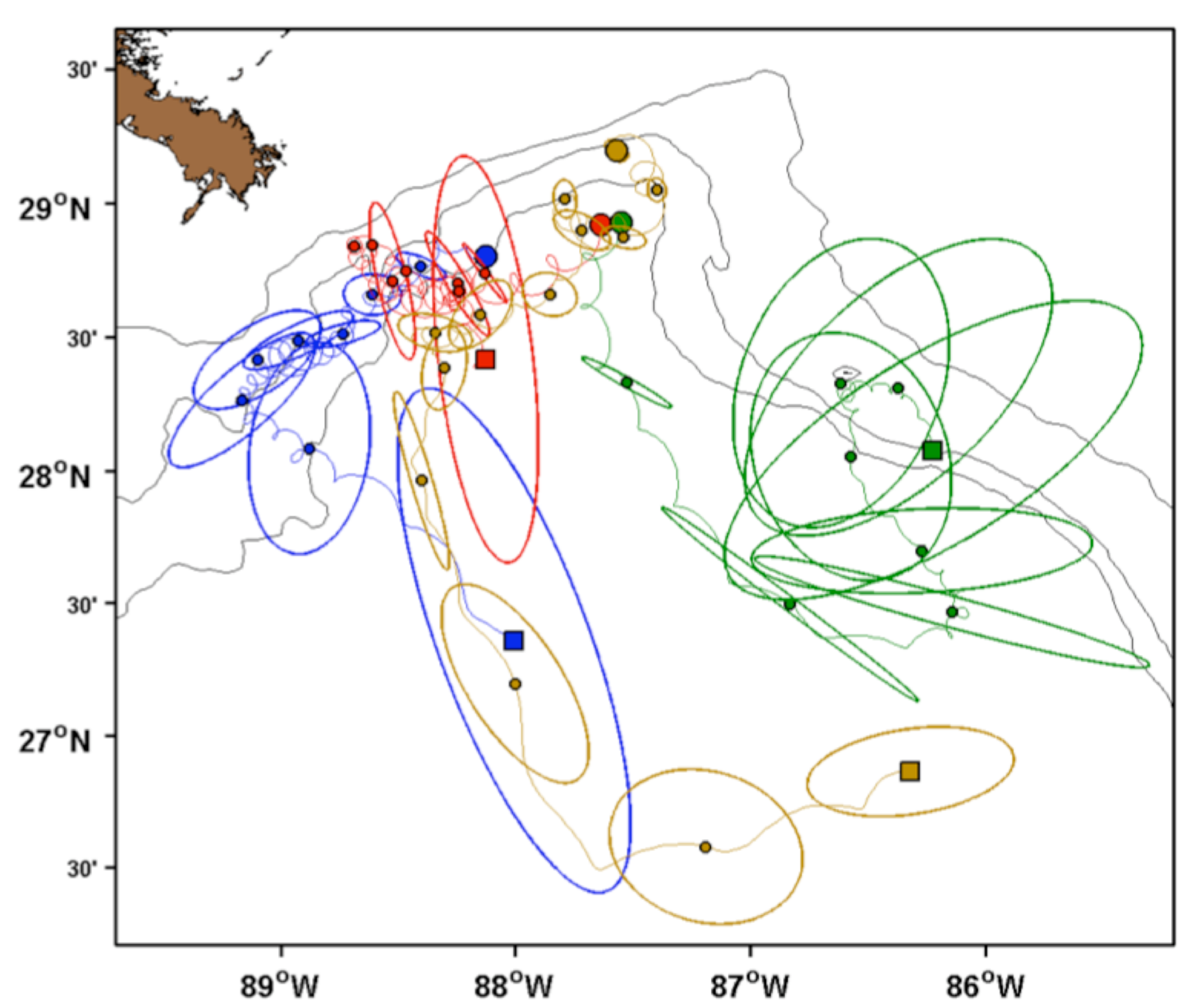}
\end{center}
\caption{{\bf Dispersion ellipses.} Trajectories and dispersion ellipses for 
$\mathcal{S}$1 (blue)  and
$\mathcal{T}$1 (yellow). 
Launch $\mathcal{S}$2 has been separated into two groups; drifters initialized in MRO water 
with residence times in the Canyon longer than 7 days(red),
and those with residence times in the Canyon less than 7 days (green).
}
\label{fig5}
\end{figure*}

Spatial and temporal distributions of basic dispersion statistics for four drifter groups 
are shown in Fig. \ \ref{fig5}. The $\mathcal{S}$2  launch has been split into two groups based on residence 
time and surface salinity characteristics: drifters launched in low surface salinity water 
with residence times greater than 7 days (referred to as MRO drifters) and those launched 
in higher surface salinity water with short residence times. Center of mass 
trajectories for each group (symbol marking every three days) as well as dispersion ellipses 
indicating the size and orientation of the standard deviation of the relative dispersion of 
drifters about the cluster center of mass are plotted. Observations confirm the disparity between 
slow, isotropic dispersion inside the canyon and rapid stretching outside.

\begin{figure*}[ht]
\begin{center}
\includegraphics[width=0.48\textwidth]{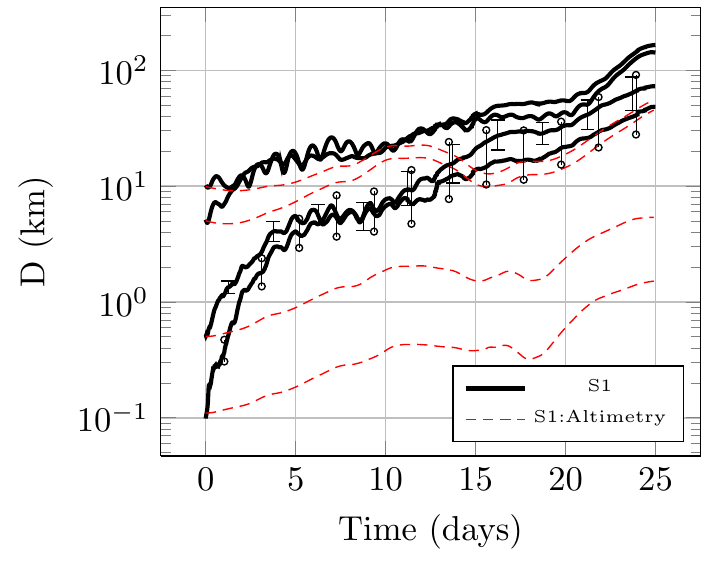}
\includegraphics[width=0.48\textwidth]{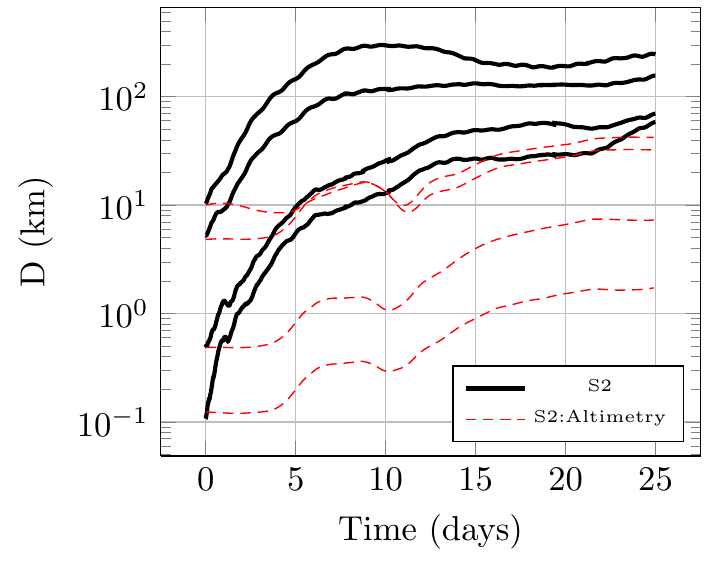} 

\medskip
\includegraphics[width=0.62\textwidth]{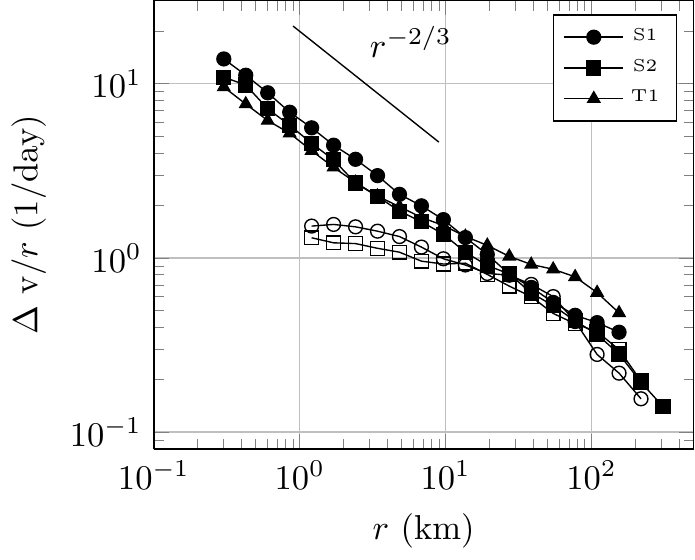}
\end{center}
\caption{{\bf Dispersion diagrams from GLAD in comparison to NCOM and AVISO.}
Time dependence of the relative dispersion, $D(t)$, for four different initial separation distances for the $\mathcal{S}$1 (top) and 
$\mathcal{S}$2 (middle) launches. 
For comparison, data from identical launches advected using geostrophic velocities produced by AVISO altimeter data are shown in red.
Lower panel:  The scale-dependent pair separation rate as function of separation distance for
the three launches ($\mathcal{S}$1, $\mathcal{S}$2, $\mathcal{T}$1) shown in solid symbols with corresponding model results from a 3-km resolution NCOM simulation for $\mathcal{S}$1 and $\mathcal{S}$2 
shown in open symbols. 
The slope indicates the Richardson regime, $\Delta v/r \sim r^{-2/3}$.}
\label{fig3}
\end{figure*}

The top panels in Fig.\ \ref{fig3} show relative dispersion curves (here $D(t)$) for $\mathcal{S}$1 and $\mathcal{S}$2
conditioned on the initial separation distance of drifter pairs.
Initial separation bins are centered at 0.1, 0.5, 5 and 10km.
Data densities range from 48 drifter pairs for the 0.1-km
$\mathcal{S}$1 bin to 1034 pairs for the 10-km $\mathcal{S}$1 bin.
\textcolor{black}{Error bars, shown for the smallest initial separations in $\mathcal{S}$1, were computed from
standard 95\% confidence intervals produced by 2,000 bootstrapped samples at each time.}
Both launches indicate that initial
growth depends strongly on the initial separation scale with faster growth rates for smaller separations.
In the $\mathcal{S}$1 launch, which was entirely confined to the canyon for
one week, dispersion from initial scales below 1km is arrested at $\sim$8-km length scales while dispersion
from initial scales above 1km
shows arrest at $\sim$30-km length scales. All curves indicate considerable energy at near-inertial frequencies.
Similar behavior is observed in the $\mathcal{S}$2 data for the smallest separation scale.
Corresponding dispersion curves derived from artificial drifters
(launched at the same initial time and position as the GLAD drifters) advected
by the geostrophic velocity field derived from AVISO gridded altimeter data do not exhibit this pattern.
\textcolor{black}{Neither relative nor absolute dispersion metrics for any of the drifter launches 
exhibited asymptotic behavior 28 days after release.}

Scale-dependent dispersion results are displayed in the bottom panel of Fig.\ \ref{fig3} where, 
for all three clusters, the dispersion rate given by the time-scale
$\lambda(r) = \Delta v(r)/r$ scales with $r^{-\beta}, \beta \ne 0$ for separation scales below 10 km. The observed exponent in each case is consistent
with Richardson's two-thirds law and a local dispersion regime where the underlying Eulerian 
kinetic energy spectrum scales is considerably shallower than the $E(k) \sim k^{-3}$ spectrum
expected in an enstrophy cascading regime.  
Comparison of dispersion rates for synthetic drifters launched at identical locations and times to those in $\mathcal{S}$1 and $\mathcal{S}$2, and advected by a data-assimilating, 
%$\Delta x = 3$ km,  
operational model (Navy Coastal Ocean Model, NCOM) simulation of the Gulf show reasonable agreement with data at mesoscales ($r > 10$km), but poor agreement 
at submesoscales where the model fields necessarily impose steep spectral decay near the model grid spacing.
\textcolor{black}
{In contrast to the situation where small-scale dispersion is dominated by the strain of large scale, nearly two-dimensional ocean
flows,  the observations clearly indicate the presence of energetic, local contributions to surface relative dispersion on scales $<10$km in the DeSoto Canyon region. As such, the observed measures at the submesoscales show considerably 
faster relative dispersion than that seen in either altimetry-derived or model-based velocity fields.}

Following Richardson's original work \textsuperscript{\cite{Richardson:1926}},
the increased rate of spread of a growing contaminantpatch can be quantified by a scale dependent dependent relative diffusivity, $K(
l)$, definedby
$\frac{\partial q}{\partial t} = \frac{\partial}{\partial l} K(l) \frac{\partial q}{\partial l}$
where $q(l,t)$ is the distribution of separation lengths, $l$. Richardson's orig
inal  empirical fit
 for atmospheric data, $K(l) \propto l^{4/3}$, was also derived by
Obukhov\textsuperscript{\cite{Obukhov41}}, from similarity theory for turbulence
 with a forward energy cascade.

\begin{figure*}[ht]
\begin{center}
\includegraphics[width=0.62\textwidth]{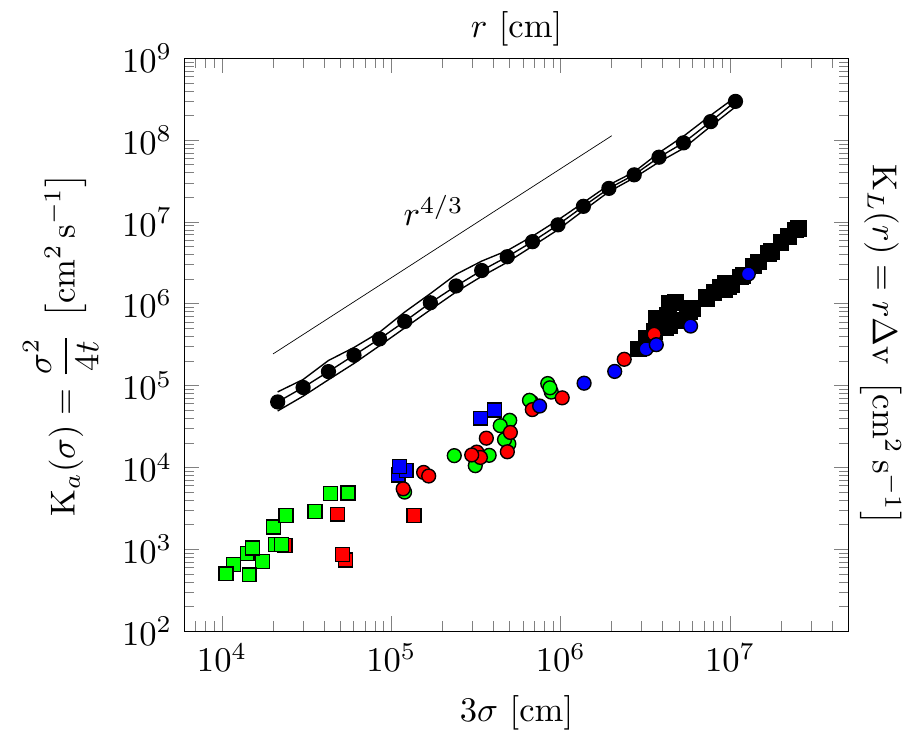}
\end{center}
\caption{{\bf Scale-dependent relative diffusivities:} 
\textcolor{black}
{
Bottom and left axes: Tracer-based diffusivity estimates based on fitting ellipses.
The solid red, green and blue symbols show
Okubo\textsuperscript{\cite{Okubo1971}} estimates of
$K_a(3 \sigma) = \sigma^2/ 4t$.
Corresponding estimates from the $\myS$ drifter data are shown
by solid black squares.
Right} and top axes: Scale dependent mixing length diffusivities,
 $K_L(r) = r \Delta v(r)$,
observed in $\myS$ launch plotted with uncertainty estimates in solid black lines and filled
black circles.
Richardson-Obukhov scaling law, $K_L(r) \sim r^{4/3}$ is indicated.
}
\label{fig4}
\end{figure*}

Dye-based observations estimate the diffusivity, $K_a(l)$, from area growth rates defined by fitting Gaussian
ellipses to the evolving concentration patch observed along ship-tracks\textsuperscript{\cite{sundermeyer2001}};
$4 K_a = d \sigma^2/dt$, where  $\sigma^2(t) = 2\, \sigma_a\,\sigma_b $ and
$(\sigma_a,\sigma_b)$ measure the major and minor axes of the ellipse.
The scale-dependent diffusivity is typically estimated by assuming a fixed value of $K_a$ and integrating
to arrive at $K_a(l) = \sigma^2(t)/4t$ where, following Okubo\textsuperscript{\cite{Okubo1971}}, the
scale length is given by $l = 3 \sigma$.
The left and lower axes of Fig.\ \ref{fig4} show the scale-dependent relative diffusivity defined this way as
calculated from the dispersion ellipses observed during the $\myS$ launch
(Fig.\ \ref{fig5}).
The single launch drifter estimates, plotted every
12 hours starting 4 days after the launch, are shown in solid black squares while the colored filled symbols
show Okubo's compilation of individual dye experiments spanning a number of oceans over the course of several years.
The magnitudes of the drifter based diffusivity values are remarkably consistent with traditional
dye-based observations and clearly extend $K_a(l) \sim l^{4/3}$
Richardson-Obukhov scaling to larger scales.

The above estimates of $K_a(\sigma)$ rely on fitting
observed distributions with Gaussian ellipses.
To investigate diffusivity scaling over the full range of
separation scales available in the drifter data, we also examine
a Lagrangian upper-bound diffusivity estimate based on classic mixing length arguments.
The line marked by solid black circles in Fig.\ \ref{fig4} shows
$K_L(r) = r \Delta v(r)$ for the $\myS$ launch.
Also shown are uncertainty estimates
given by the 95\% confidence intervals produced by 500 bootstrapped samples of 200 randomly chosen times during the 28 day launch.
The observations are  well fit by classical Richardson-Obukhov scaling over the entire 200m -- 100km range of
 available separation scales.

\section{Conclusions}
Large numbers of accurate, high-frequency Lagrangian instruments, launched nearly simultaneously
provide an effective means for quantifying scale-dependent dispersion at the ocean's surface. 
In the DeSoto Canyon region, an energetic submesoscale field clearly produces local dispersion at $\sim$100-m scales
which is not captured by ocean surface velocity fields derived from current satellite altimetry or operational ocean models.
The high energy of the observed submesoscale field has significant implications for both predictive 
modeling of oceanic pollutant discharges in this region as well as for understanding overall mechanisms for 
energy transfer in the ocean.
Whether the predominence of submesoscale fluctuations in setting local dispersion properties is
an inherent surface feature of the global ocean or is instead a confined result due the complexities of local forcing 
mechanisms in the DeSoto Canyon region,  
can be directly  addressed by conducting similar large-scale, synoptic Lagrangian observational programs in other locations.

\section{Materials}
The GLAD program was based on 300 GPS-equipped CODE drifters with nominal position accuracy of 5 m
and battery life exceeding two months
to provide a new level of statistical accuracy in measuring two-point Lagrangian velocity and displacement statistics.
A special agreement was made with the Globalstar company to retrieve the positions at 5-min intervals.
Post-processing consisted of identifying inconsistent short-term position sequences near large reception gaps and
the removal of outliers in position or velocity magnitude and rotation.
Data gaps were filled using a non-causal spline interpolation, and the clean drifter position data was then low-pass filtered with a 1-h cut-off
and resampled at uniform 15-min intervals.

Along-track salinity was collected with an on-board flow-through system
using a Seabird thermosalinograph (SBEMicroTSG45) and external temperature
sensor (SBETemp38) located approximately 2 m below the water level at the ship's bow. Salinity,
expressed in PSU calculated using the PSS equations, has an estimated initial accuracy of 0.005 PSU and monthly drift of 
$<$0.003 PSU. Simultaneous salinity and 3-m accurate ship position (via an onboard Furuno GP90) were logged at 5-s intervals. 

The geostrophic velocity field is assumed to be of the form $
   v(x,t) = gf(x_2)^{-1}\nabla^\perp\eta(x,t) +
   \nabla\varphi(x,t),
   \label{eq:v}
$
where $g$ is the acceleration of gravity; $f(x_2)$ is the
latitude-dependent Coriolis parameter; $\eta(x,t)$ is sea surface height
anomaly from AVISO and
$\varphi(x,t)$ is such that $v(x,t)$ is non-divergent in the interior and its normal projection  at
the coastline vanishes.
The
steady  $\eta(x,t)$ component is given by a mean dynamic
topography constructed from altimetry data, {\em in situ} measurements,
and a geoid model
while the transient
component is given by gridded altimetric SSH anomaly 
measurements provided weekly at 0.25$^{\circ}$-resolution
Jason-1 and 2, and Cryosat-2 traversed the 
Gulf of Mexico about 10 times  per week on average during the study period.

\begin{acknowledgments}
This research was made possible by a grant from 
BP/The Gulf of Mexico Research Initiative. 
\end{acknowledgments}

\bibliographystyle{plain}
%\bibliography{mybibs}

\end{document}